\documentclass[12pt]{article}

\usepackage{graphicx,amssymb,url,mathrsfs, amsmath}
\usepackage{eucal}
\usepackage{wrapfig}
\usepackage{boxedminipage}
\usepackage{setspace}
\usepackage{subfigure}

\def\a  {\alpha}                
       \def\d  {\delta}        
\def\e  {\epsilon}        \def\k  {\kappa}
\def\l  {\lambda}             \def\m  {\mu}
          \def\s  {\sigma}        
\def\t  {\tau}

\newcommand{\cala}{\mbox{${\cal A}$}} 
 
\newcommand{\cale}{\mbox{${\cal E}$}}

\newcommand{\calk}{\mbox{${\cal K}$}} \newcommand{\call}{\mbox{${\cal L}$}}
 
\newcommand{\calo}{\mbox{${\cal O}$}} 
 \newcommand{\calr}{\mbox{${\cal R}$}}

\def\IR{{\hbox{{\rm I}\kern-.2em\hbox{\rm R}}}}
\def\IB{{\hbox{{\rm I}\kern-.2em\hbox{\rm B}}}}
\def\IN{{\hbox{{\rm I}\kern-.2em\hbox{\rm N}}}}
\def\IC{\,\,{\hbox{{\rm I}\kern-.59em\hbox{\bf C}}}}
\def\IZ{{\hbox{{\rm Z}\kern-.4em\hbox{\rm Z}}}}
\def\IP{{\hbox{{\rm I}\kern-.2em\hbox{\rm P}}}}
\def\IH{{\hbox{{\rm I}\kern-.4em\hbox{\rm H}}}}
\def\ID{{\hbox{{\rm I}\kern-.2em\hbox{\rm D}}}}

\def\be{\begin{equation}}
\def\ee{\end{equation}}
\def\ba{\begin{eqnarray}}
\def\ea{\end{eqnarray}}

\def\half{\frac{1}{2}}

\newcommand{\inv}[1]{\frac{1}{#1}}

\def\ra{\rightarrow}

\def\dell{\partial}

\def\Tr{{\rm tr}\,}

\def\det{{\rm det}}

\def\nn{\nonumber}
\def\ea{{\it et al}. }

\def\De{\textrm{D8}}
\def\DeB{\overline{\textrm{D8}}}

\newcommand{\Ukk}{U_{\rm KK}}
\newcommand{\Mkk}{M_{\rm KK}}
\newcommand{\wt}{\widetilde}
\newcommand{\wh}{\widehat}

\newcommand{\Az}{\cala_{0}}

\begin{document}

\begin{titlepage}

%\begin{flushright}
%  {\tt hep-th/0608046}
%\end{flushright}
%\vspace{0.5in}

\begin{center}
  {\large \bf Dense Holographic QCD in the Wigner-Seitz Approximation} \\
\vspace{10mm}
  Keun-Young Kim$^a$, Sang-Jin Sin $^{a,b}$  and Ismail Zahed$^a$\\
\vspace{5mm}
$^a$ {\it Department of Physics and Astronomy, SUNY Stony-Brook, NY 11794}\\
$^b$ {\it Department of Physics, BK21 division, Hanyang University, Seoul 133-791, Korea}\\
\vspace{10mm}
% {\tt \today}
\end{center}
\begin{abstract}
We investigate cold dense matter in the context of Sakai and Sugimoto's holographic
model of QCD  in the Wigner-Seitz approximation. In bulk, baryons are treated as instantons on $S^3\times R^1$ in
each Wigner-Seitz cell.  In holographic QCD,  Skyrmions are instanton
holonomies along the conformal direction. The high density phase is identified with
a crystal of holographic Skyrmions with restored chiral symmetry at about $4\Mkk^3/\pi^5$.
As the average density goes up, it approaches to uniform distribution while the chiral condensate approaches to  p-wave  over a cell.  The chiral symmetry is effectively restored in
long wavelength limit since  the chiral order parameter is
averaged to be zero over a cell. The  energy density in dense medium varies as  $n_B^{5/3}$, which is the expected power for  non-relativistic fermion.
This shows that the Pauli exclusion  effect in boundary  is   encoded in the Coulomb repulsion in the bulk.
\end{abstract}

\end{titlepage}

\renewcommand{\thefootnote}{\arabic{footnote}}
\setcounter{footnote}{0}

%\tableofcontents
\newpage

%%%%%%%%%%%%%%%%%%%%%%%%%%%%%%%%%%%%%%%%%%%%%%%%%%%%%%%%%%%%%%%%%%%%%%%%%%%
%%%%%%%%%%%%%%%%%%%%%%%%%%%%%%%%%%%%%%%%%%%%%%%%%%%%%%%%%%%%%%%%%%%%%%%%%%%
%%%%%%%%%%%%%%%                 BODY                    %%%%%%%%%%%%%%%%%%%
%%%%%%%%%%%%%%%%%%%%%%%%%%%%%%%%%%%%%%%%%%%%%%%%%%%%%%%%%%%%%%%%%%%%%%%%%%%
%%%%%%%%%%%%%%%%%%%%%%%%%%%%%%%%%%%%%%%%%%%%%%%%%%%%%%%%%%%%%%%%%%%%%%%%%%%

\section{Introduction}

The AdS/CFT approach~\cite{Maldacena} provides a powerful
framework for discussing large $N_c$ gauge theories at strong
coupling $\lambda=g^2N_c$. The model suggested by Sakai and
Sugimoto (SS) \cite{Sakai1} offers a specific holographic
realization that includes $N_f$ flavors and is chiral. For
$N_f\ll N_c$, chiral QCD is obtained as a gravity dual to $N_f$
$\De$-$\DeB$ branes embedded into a  D4 background in 10
dimensions where supersymmetry is broken by the Kaluza-Klein (KK)
mechanism. The SS model yields a holographic description of
pions, vectors, axials and baryons that is in good agreement with
experiment~\cite{Sakai1, Sakai3, Baryons}. The SS model at finite
temperature has been discussed in~\cite{FiniteT} and at finite
baryon density in~\cite{Finitemu,Rozali,KSZ2}.
The finite baryon chemical potential problem in D3/D7 model was discussed in \cite{D3D7} and
 Isospin chemical potential and glueball decay also has been discussed in~\cite{Others}.

Cold and dense hadronic matter in QCD is difficult to track from
first principles in current lattice simulations owing to the sign
problem. In large $N_c$ QCD baryons are solitons and a dense
matter description using Skyrme's chiral model~\cite{Skyrme} was
originally suggested by Skyrme and others~\cite{DENSESKYRME}. At
large $N_c$ and high density matter consisting of solitons
crystallizes, as the ratio of potential to kinetic energy
$\Gamma=V/K\approx N_c^2$ is much larger than 1. QCD matter
at large $N_c$ was recently revisited in~\cite{ROBE}.

The many-soliton problem can be simplified in the crystal limit by
first considering all solitons to be the same and second by reducing
the crystal to a single cell with boundary conditions much like the
Wigner-Seitz approximation in the theory of solids. A natural way to
describe the crystal topology is through $T^3$ with periodic boundary
conditions. In so far, this problem can only be addressed numerically.
A much simpler and analytically tractable approximation consists of
treating each Wigner-Seitz cell as $S^3$ with no boundary condition
involved. The result is dense Skyrmion matter on $S^3$ \cite{DENSES3}.
Interestingly enough, the energetics of this phase is only few percent
above the energetics of a more involved numerical analysis based on $T^3$.
Skyrmions on $S^3$ restore chiral symmetry on the average above a
critical density. While Skyrmions on $S^3$ are unstable against $T^3$,
they still capture the essentials of dense matter and chiral restoration
in an analytically tractable framework.

Cold dense matter in holographic QCD is a crystal of instantons
with $\Gamma=\sqrt{\lambda}/v_F\gg 1$ where $v_F\approx 1/N_c$ is
the Fermi velocity. (In contrast hot holographic QCD has
$\Gamma=\sqrt{\lambda}\gg 1$). When the wigner-Seitz cell is {\it
approximated} by $S^3$, the pertinent instanton is defined on
$S^3\times R$. In this paper, we investigate cold QCD matter
using instantons on $S^3\times R$ in bulk. As a result the
initial D4 background is {\it deformed} to accomodate for the
$S^3$ which is just the back reaction of the flavour crystal
structure on the pure gauge theory. Holographic dense matter can
be organized in $1/\lambda$ at large $N_c$.
In our model the baryon density is uniform while the chiral condensate  is p-wave  over a cell.
However, the chiral condensation is
averaged to be zero over a cell so that  the chiral symmetry is effectively restored in
long wavelength limit.
We will show that as the average density goes up, it approaches to uniform distribution while the chiral condensate approaches to  p-wave  over a cell.   The  energy density in dense medium varies as  $n_B^{5/3}$, which is the expected power for  non-relativistic fermion.
This shows that the Pauli exclusion  effect in boundary  is   encoded in the Coulomb repulsion in the bulk.

In section 2, we
define this deformation and discuss the D8 brane embedding
structure. The instantons on $S^3\times R$ in the flavour D8
brane is discussed in section 3. In section 4,5,6 we derive the
equation of state of cold holographic matter using the small size
instanton expansion and in general. In section 7 we show how the
holographic small instantons in bulk transmute large size
Skyrmions on the boundary. The comparison to other models of
nuclear matter is carried in section 8. Our conclusions are in
section 9.

\section{D8 brane action}

We consider crystallized skyrmions at finite density in the
Wigner Seitz approximation. Spatial $R^3$ is naturally converted
to $T^3$ with periodic boundary conditions. As a result the D4
background geometry is deformed. The baryons are then instantons
on $T^3\times R$. Most solutions are only known numerically on
the lattice. A simpler and analytically tractable analysis that
captures the essentials of dense matter is to substitute $T^3$ by
$S^3$ in bulk with no boundary conditions altogether. As a
result, the D4 background dual to the crystal is modified with
the boundary special space as $S^3$. Specifically, the 10
dimensional space is that of $ (R^1 \times S^3) \times R^1 \times
S^4$. The ensuing metric on  D4 is therefore
\begin{eqnarray}
&&ds^2=\left(\frac{U}{R}\right)^{3/2} \left(-dt^2 + \calr^2 d
\Omega_3^2  +f(U)d\tau^2\right) +\left(\frac{R}{U}\right)^{3/2}
\left(\frac{dU^2}{f(U)}+U^2 d\Omega_4^2\right)\  , \label{D4metric} \\
&&d\Omega_3 \equiv d\psi^2 + \sin^2\!\psi\ d\theta^2 +
\sin^2\!\psi \sin^2\!\theta \ d\phi^2\ , \quad f(U)\equiv
1-\frac{\Ukk^3}{U^3} \ , \\
&&e^\phi= g_s
\left(\frac{U}{R}\right)^{3/4}, ~~F_4\equiv dC_3=\frac{2\pi
N_c}{V_4}\epsilon_4 \ , ~~~ \label{D4sol}
\end{eqnarray}
While this compactified  metric is not an exact solution to the
general relativity (GR) equations for small size $S^3$, it can be
regarded as an approximate solution for large size $S^3$. Indeed,
in this case, the GR equations are seen to be sourced by terms
wich are down by the size of $S^3$. Here, (\ref{D4sol}) can be
regarded as an approximation to the stable metric with $T^3$
for a dense matter analysis. Clearly, the former is unstable
against decay to the latter, which will be reflected by the
fact that the energy of dense matter on $S^3$ is higher than
that on $T^3$. As indicated in the introduction, the Skyrmion
analysis shows that the energy on $S^3$ is only few percent
that of $T^3$. So we expect the current approximation to capture
the essentials of dense matter in holographic QCD. Specifically,
the nature and strength of the attraction and repulsion in dense
matter. Indeed, this will be the case as we will detail below.

Now, consider $N_f$ probe D8-branes in the $N_c$ D4-branes
background. With $U(N_f)$ gauge field $A_M$ on the D8-branes, the
effective action consists of the DBI action and the Chern-Simons
action
\begin{eqnarray}
S_{\De}&=& S_{\mathrm{DBI}} + S_{\mathrm{CS}} \ , \label{Action.0} \nn \\
S_{\mathrm{DBI}}&=& -T_8 \int d^9 x \ e^{-\phi}\ \Tr
\sqrt{-\det(g_{MN}+2\pi\alpha' F_{MN})} \ , \label{DBI}\\
S_{\mathrm{CS}}&=&\frac{1}{48\pi^3} \int_{D8} C_3 \Tr F^3
\label{CS} \ .
\end{eqnarray}
where $T_8 = 1/ ((2\pi)^8 l_s^9)$, the tension of the D8-brane,
$F_{MN}=\partial_M A_N -\partial_N A_M -i \left[ A_M , A_N
\right]$ ($M,N = 0,1,\cdots,8$), and $g_{MN}$ is the induced
metric on  the D8-branes
\begin{eqnarray}
&& ds^2_{D8}=\left(\frac{U}{R}\right)^{3/2}\!\!\!\!  (-dt^2 +
\calr^2 d \Omega_3^2) + g_{\s\s}d\s^2+
\left(\frac{R}{U}\right)^{3/2}\!\!\!\! U^2 d\Omega_4^2 \ , \label{D8metric} \\
&& g_{\s\s} \equiv G_{\t\t}\dell_\s\t\dell_\s\t +
G_{UU}\dell_{\s}U\dell_{\s}U \ ,
\end{eqnarray}
where $G_{MN }$ refer to the background metric (\ref{D4metric})
and the profile of the D8 brane is parameterized by $U(\s)$ and
$\t(\s)$.

The gauge field $A_M$ has nine components, $\Az$, $A_i =
A_{1,2,3}$, $A_\s(=A_4)$, and $A_{\a }$($\a = 5,6,7,8$, the
coordinates on the $S^4$). We assume
\begin{eqnarray}
&& \Az = \Az(\s) \in U(1) \ , \\
&& (A_i = A_i(x^i, \s), \quad A_\s = A_\s(x^i, \s)) \in SU(N_f) \ , \\
&& A_\a = 0 \ .
\end{eqnarray}
Then the action becomes 5-dimensional:
\begin{eqnarray}
&&S_{\mathrm{DBI}}= -\frac{8 \pi^2 T_8 R^3 }{3 g_s}\ \Tr\int dt \e_3 d\s \ U \nn \\
&&\quad \left[\quad \left\{ \left(\frac{U}{R}\right)^{3/2}g_{\s\s}
- (2\pi\a')^2 ( \dell_\s \Az)^2 \right\}\left\{
\left(\frac{U}{R}\right)^{3}
+\half (2\pi\a')^2 F_{ij}F^{ij} \right\} \right. \nn \\
&&\quad \quad \left. + \ \left(\frac{U}{R}\right)^{3} (2\pi\a')^2
F_{\s i}F_{\s} ^{\ i} + \frac{1}{4}(2\pi\a')^4 (\e_{ijk}
F_{i\s}F_{jk})(\e_{ijk} F^{i}_{\ \s }F^{jk}) \quad \right]^{1/2}
\ \label{DBI.1}
, \\
&&S_{\mathrm{CS}} = \frac{N_c}{24\pi^2} \Tr \int \cala \wedge F
\wedge F \ ,
\end{eqnarray}
where $\e_3$ is the volume form of $S^3$ space and the indices
$i,j,k ( \in \{\psi, \theta, \phi\})$ are raised by the metric
$\tilde{g}^{ij}$ defined by
\begin{eqnarray}
\tilde{g}^{ij} = \left(\inv{\calr^2} , \inv{\calr^2 \sin^2\psi} ,
\inv{\calr^2 \sin^2\psi \sin^2\theta} \right) \ .
\end{eqnarray}

\section{Instanton in $S^3 \times R^1$}

Only $\Az$ will be determined dynamically in the given instanton
background $A_i,A_\s$. The exact background instanton solution is
unknown. Thus we start with an approximate solution which is the
SU(2) Yang-Mills instanton solution in the space with metric,
\begin{eqnarray}
ds^2 = d\s^2 + \calr^2d\Omega_3 \ . \label{RWmetric}
\end{eqnarray}
This metric is different from our metric in (\ref{D8metric}) and
(\ref{DBI.1}), where there are warping factors. Furthermore our
action is the nonlinear DBI action and not a Yang-Mills action.
However it can be shown that the Yang-Mills instanton in the space
(\ref{RWmetric}) is the leading order solution of $1/\l$
expansion of the full metric and the DBI action as shown in
\cite{Sakai3} . So the solution can be used in the leading order
calculation.

We summarize here the (anti) self dual instanton solution
obatained in \cite{Forkel}. Using the ansatz ,
\begin{eqnarray}
A = f(\s)U^{-1}dU \ , \quad \quad U \equiv \cos \psi + i \t_a
\hat{r}^a(\theta, \phi)\sin \psi \ ,
\label{INS3}
\end{eqnarray}
we get the field strength, in terms of vielbein whose relation to
the coordinate $\psi,\theta, \phi,$ is specified in \cite{Forkel},
\begin{eqnarray}
&&F =  \frac{(\dell_\s f) \t_a}{\calr} \ e^0 \wedge e^a + \half
\left[ \frac{2(f^2-f)\t_d\epsilon^{d}_{\ bc}}{\calr^2} \right] \
e^b \wedge e^c \ , \label{Field.strength}
\end{eqnarray}
where we used $L_a = U^{-1}\dell_a U = \t_a/\calr$.
If we require (anti) self-duality,
\begin{eqnarray}
\dell_\s f = \pm \frac{2(f^2 - f)}{\calr} \ ,
\end{eqnarray}
then $f$ is determined as
\begin{eqnarray}
&&f_{\pm} \equiv \inv{1+e^{\mp2(\s-\s_0)/\calr}}\ ,
\end{eqnarray}
so the field strength of one (anti) instanton solution is
\begin{eqnarray}
&&F^{\pm} = (\dell_\s f_{\pm})\frac{\t_a}{\calr} (e^0 \wedge e^a
\pm \half \epsilon^{a}_{\ bc} e^b \wedge e^c ) \ .
\end{eqnarray}

\section{D8 brane plus Instanton}\label{Sec.sol}

Now that we have the background instanton configurations, the
remaining dynamical variables are $\t$ and $\Az$. However it can
be shown that $\dell_\s \t(\s) = 0$ is always a solution of the
Euler-Lagrange equation regardless of the gauge field. For
simplicity we will work with this specific configuration so that
the only dynamical variable is $\Az$.
Let us parameterize $\Az$ by $Z$ defined as
\begin{eqnarray}
&& U \equiv (\Ukk^3 + \Ukk \s^2)^{1/3} \ , \nn \\
&& Z \equiv \frac{\s}{\Ukk} \ , \quad K \equiv 1+Z^2 \ .
\end{eqnarray}
Then the field strength is expressed in terms of Z and the
dimensionless radius $\wh{\calr} \equiv \calr / \Ukk$,
\begin{eqnarray}
&&F_{Za} = \inv{\Ukk} f'\ \frac{\t_a}{\wh{\calr}  } \ , \nn \\
&&F_{ab} = \inv{\Ukk^{2}} f' \ \frac{\epsilon_{ab}^{\ \
c}\t_c}{\wh{\calr} } \ , \label{Field.strength.1}
\end{eqnarray}
where $f' \equiv \dell_Z f$. The instanton configuration is
\begin{eqnarray}
f_{\pm} = \inv{1+e^{\mp2(Z-Z_0)/\wh{\calr}}} \ . \label{Instanton}
\end{eqnarray}

The DBI action reads
\begin{eqnarray}
&&S_{\mathrm{DBI}}= - \frac{8 \pi^2 T_8 R^3 }{3 g_s}\ \Tr\int dt \e_3 d Z K^{1/3}   \nn \\
&&\quad \left[\quad \left\{
\left(\frac{4}{9}\right)\Ukk^2K^{-1/3} - (2\pi\a')^2 (\Az')^2
\right\}\left\{ K\left(\frac{\Ukk}{R}\right)^{3}
+\half (2\pi\a')^2 F_{ab}^2 \right\} \right. \nn \\
&&\quad \quad \left. + \ K\left(\frac{\Ukk}{R}\right)^{3}
(2\pi\a')^2 F_{Z a}^2 + \frac{1}{4}(2\pi\a')^4 (\e_{abc}
F_{aZ}F_{bc})^2 \quad \right]^{1/2} \ ,
\end{eqnarray}
where $\Az' \equiv \dell_Z \Az$. We are using the same vielbein
coordinates as (\ref{Field.strength}). Since the instanton size
(${\calr}$) is of order $\calo(\l^{-1/2})$ we define a new
dimensionless parameter $\wt{\calr}$, which is order of $(\l^0)$,
as
\begin{eqnarray}
\wt{\calr} \equiv \sqrt{\l} \wh{\calr} =
\sqrt{\l}\frac{\calr}{\Ukk} \ . \label{Rtilde}
\end{eqnarray}
Furthermore we rescale the coordinate and the instanton field
strength for a systematic $1/\l$ expansion
\begin{eqnarray}
&&x^a \ra \l^{-1/2}x^a \ , \quad Z \ra \l^{-1/2}Z \  , \quad t
\ra t
\ ,  \nn \\
&& F_{ab} \ra \l F_{ab} \ , \quad F_{aZ} \ra \l F_{aZ} \ , \quad
\Az \ra \Az \ , \nn \\
&& K = (1+Z^2) \ra \left(1+ \inv \l Z^2\right) \equiv K_\l ,
\label{Rescale}
\end{eqnarray}
so all coordinates and gauge fields become of order of
$\calo(\l^0)$ and we
By using the instanton solution (\ref{Instanton}) we get
\begin{eqnarray}
&&S_{\mathrm{DBI}}= -  \frac{N_c \l}{3^9 \pi^5 \Ukk \Mkk^{-3}}  \ \Tr\int dt \e_3 d Z K_\l^{1/3}  \nn \\
&&\quad \left[\quad \left\{ 1 +  \frac{3^7 \pi^2}{4 \Mkk^2
\Ukk^2} K_\l^{1/3} \frac{f'^2}{\wt{\calr}^2} - \inv{\l}\frac{3^6
\pi^2}{4 \Mkk^2}K_\l^{1/3} ( \Az')^2
\right\} \right. \nn \\
&&\quad \quad \  \left. \left\{ \Mkk^2\Ukk^2K_\l^{4/3} + \frac{3^7
\pi^2}{4 \Mkk^2 \Ukk^2} K_\l^{1/3} \frac{f'^2}{\wt{\calr}^2}
\right\}       \quad \right]^{1/2}
\end{eqnarray}
If we let $\Ukk = \Mkk^{-1}$ for simplicity, then the DBI action
yields
\begin{eqnarray}
S_{\mathrm{DBI}} &=& -d N_c \l \ \int dt \e_3 d Z \ \sqrt{\left\{
1 + K_\l^{1/3} {\wt{F}}^2 - \inv{\l} K_\l^{1/3} (\wt{\cala}_0')^2
\right\} \left\{K_\l^{3/4} + K_\l^{1/3}
{\wt{F}}^2 \right\}   } \nn \\
&=&-d N_c \l \ \int dt \e_3 d Z \ \left[ 1 + \frac{3Z^2}{8\l} +
{\wt{F}}^2 + \frac{Z^2}{3\l}{\wt{F}}^2 - \frac{1}{2\l}
(\wt{\cala}_0')^2 + \calo((1/\l)^2)\right]  \ , \label{DBI.2}
\end{eqnarray}
where
\begin{eqnarray}
&&d \equiv \frac{ 2 \Mkk^4 }{3^9 \pi^5} \ ,  \quad  \wt{\cala}_0
\equiv \frac{3^3 \pi }{2 \Mkk}\Az \ , \quad {\wt{F}}^2 \equiv \frac{3^7 \pi^2}{4} J\ , \nn \\
&& J\equiv \frac{f'^2}{\wt{\calr}^2} =
\frac{\mathrm{sech}^4(Z/\wt{\calr})} {4 \wt{\calr}^4 }  \sim
\frac{1}{3 \wt{\calr}^3} \d (Z) \ , \nn \\
&& \dell_Z \calk \equiv \dell_Z \frac{1}{6\wt{\calr}^3}
\left[\tanh(Z/\wt{\calr})
\left(1+\half\mathrm{sech}^2(Z/\wt{\calr})\right)\right] \sim
\frac{1}{6 \wt{\calr}^3}\mathrm{sgn}(Z) \label{Def.variables}
\end{eqnarray}

%\begin{eqnarray}
%S_{\mathrm{CS}} &=&  \frac{N_c}{8\pi^2} \Tr\!\!\int dt dZ
%d\Omega_3 \Az \left( \frac{24}{\Ukk^3} \frac{2
%f'(f^2-f)}{\wt{\calr}^3}\right)
%\end{eqnarray}
The Chern-Simons action does not change by the recaling
(\ref{Rescale}) , and it is order of $\l^0$. With the instanton
solution (\ref{Instanton}) the Chern-Simons action reduces to
\begin{eqnarray}
S_{\mathrm{CS}} &=& \frac{N_c}{24\pi^2} \Tr \int \cala \wedge F
\wedge F =  \frac{N_c}{8\pi^2} \Tr\!\!\int dt \e_3 dZ \Az
\inv{2}\left( 24 \Mkk^3 \frac{ f'^2}{\wt{\calr}^2}\right) \nn
\\ &=&  c N_c \int dt \e_3 dZ \wt{\cala}_0 {\wt{F}}^2   \ ,
\label{CS.2}
\end{eqnarray}
where
\begin{eqnarray}
c \equiv \frac{4 \Mkk^4}{3^9 \pi^5} \ .
\end{eqnarray}
It also can be written as
\begin{eqnarray}
S_{\mathrm{CS}} = 3N_c \wt{\calr}^3 \int dZ \Az \dell_Z \calk \ra
N_c \ , \quad \ \mathrm{\ for \ } \Az = 1 \ ,
\end{eqnarray}
which confirms that the field configuration
(\ref{Field.strength.1}), and (\ref{Instanton}) describe the
single (anti) self dual instanton since $S_\mathrm{CS}$
corresponds to $N_c$ $\times$ the Pontryagin index when $\Az = 1$.

\section{Equation of State in $1/\lambda$}

The equation of state of cold holographic matter is the energy following
from the action functional. The total action up to order of $\l^0$ is
\begin{eqnarray}
S &\equiv& \int dt \e_3 d Z ( \call_{\mathrm{DBI}} + \call_{\mathrm{CS}}) \nn \\
&=&-d N_c \int dt \e_3 d Z \ \left[\l {\wt{F}}^2 +
\frac{Z^2}{3}{\wt{F}}^2 - \frac{1}{2} (\wt{\cala}_0')^2 \right] +
c N_c \int d^4x dZ \wt{\cala}_0 {\wt{F}}^2  \ .
\end{eqnarray}
where $\Az$ is an auxillary field with no time-dependence that can
be eliminated by the equation of motion or Gauss law,
\begin{eqnarray}
\Pi' = cN_c {\wt{F}}^2 \ ,
\end{eqnarray}
with
\begin{eqnarray}
\Pi \equiv \frac{\dell \call}{\dell \wt{\cala}_0' } = d N_c
\wt{\cala}_0' \ ,
\end{eqnarray}
The integral of the equation of motion with ${\wt{F}}^2 $ in
(\ref{Def.variables}) is
\begin{eqnarray}
&& \Pi(Z) = \Pi(\infty)\left[ \tanh(Z/\wt{\calr})
\left(1+\half\mathrm{sech}^2(Z/\wt{\calr})\right)\right]  \ , \nn
\\
&&\Pi(\infty) = \frac{ \Mkk^4}{54 \pi^3}\frac{N_c}{\wt{\calr}^3} \
,
\end{eqnarray}
where we have set $\Pi(0) = 0$.

The energy of one cell is
\begin{eqnarray}
\cale_{\mathrm{cell}} &=& - \int \e_3dZ ( \call_{\mathrm{DBI}}+
\call_{\mathrm{CS}})  \nn
\\
&=& d N_c \int \e_3\  d Z \ \left[\l {\wt{F}}^2 +
\frac{Z^2}{3}{\wt{F}}^2 - \frac{1}{2}\frac{\Pi^2}{(dN_c)^2}
\right] - \int \e_3\ dZ \wt{\cala}_0 \Pi'  \ , \nn \\
&=& d N_c \ \int \e_3\  d Z \ \left[\l {\wt{F}}^2 +
\frac{Z^2}{3}{\wt{F}}^2 + \frac{1}{2}\frac{\Pi^2}{(dN_c)^2}
\right] -  \int \e_3\ \wt{\cala}_0 (Z)
\Pi(Z)\Big|_{-\infty}^{\infty} \ .
\end{eqnarray}
Thus the energy density ($\varepsilon$) of the crystalline structure is
\begin{eqnarray}
\varepsilon &\equiv& \frac{N \cale_{\mathrm{cell}} }{V} \approx
\frac{\cale_{\mathrm{cell}}}{ \int \e_3 } \nn \\
&=& d N_c \int d Z \ \left[\l {\wt{F}}^2 + \frac{Z^2}{3}{\wt{F}}^2
+ \frac{1}{2}\frac{\Pi^2}{(dN_c)^2} \right] -  \wt{\cala}_0 (Z)
\Pi(Z)\Big|_{-\infty}^{\infty} \ , \label{omega}
\end{eqnarray}
where $N$ is the total number of baryons(cells) and $V$ is the
total volume which is approximated by $N \int \e_3$.
Interestingly the second
term in (\ref{omega}) is equal to $\mu_B n_B$ since the density and the baryon chemical potential is given by
\begin{eqnarray}
n_B = \frac{1}{\int \e_3} = \frac{1}{2\pi^2 (\Ukk \wt{\calr})^3}
= \frac{1}{2\pi^2 (\sqrt{\l} {\calr})^3}\ , \qquad \m_B \equiv N_c
\Az(\infty) \ .
\end{eqnarray}
respectively and because
\begin{eqnarray}
 \wt{\cala}_0 (Z) \Pi(Z)\Big|_{-\infty}^{\infty}  = 2 \wt{\cala}_0 (\infty)
 \Pi(\infty)= N_c \Az(\infty) \frac{1}{2 \pi^2 (\Ukk \wt{\calr})^3} = \m_B n_B \ ,
\end{eqnarray}
Notice that $\wt{\calr}$ is of order of $(\l)^0$ from
(\ref{Rtilde}) so the baryon density is of order $(N_c\lambda)^0$.
%
%
% ??? Since $n_B$ is fixed we may set $\mu_B = 0$.
%
Since the action is finite and concentrated in a finite size,
we can restrict the integral to the region $Z \le Z_c$ and expand the action in $1/\lambda$.
\begin{eqnarray}
\varepsilon &=&  d N_c \int_0^{Z_c} d Z \ \left[\l {\wt{F}}^2 +
\frac{Z^2}{3}{\wt{F}}^2 + \frac{1}{2}\frac{\Pi^2}{(dN_c)^2}
\right] \label{Energy} \\
&=&  M_0 \left[n_B + \frac{a}{\l}n_B^{1/3} + \frac{b}{\l}
Z_c\,n_B^{2} \right] \ ,
\label{eos}
\end{eqnarray}
where
\begin{eqnarray}
&& M_0 \equiv 8\pi^2\k \Mkk \ , \quad \k \equiv \frac{\l
N_c}{216\pi^3}\ , \nn \\ && a  \equiv \frac{(\pi^2 - 6)\Mkk^2}{36
(2\pi^2)^{2/3}} \ , \quad b   \equiv \frac{3^6\pi^4}{2 \Mkk^3} \ ,
%\frac{Z_c}{\wt{\calr}} \ ,
\end{eqnarray}
A few remarks are in order.
\begin{enumerate}
  \item $ Z_c$ is introduced as an arbitrary cut-off which is bigger than the instanton size. However in section 8, we will argue that $ Z_c$ should be identified as
      baryon size by explicitly contructing the Skyrmion out  of the instanton
       %through Atiyah-Manton construction.
        Therefore it is not an arbitrary  number.
 \item Even in the case the instanton size is small,  the baryon size on the boundary is
not. It is of order $(N_c\lambda)^0$ and large in units of $M_{KK}$. This point
is important. While the instanton size in bulk is of the
order of the string length and thus small as $1/\sqrt{\lambda}$
in units of $M_{KK}$, its image on the boundary is a large Skyrmion.
 \item The position of the instanton $Z_0$ in the conformal direction is
set to zero by parity.
\end{enumerate}
%In~Fig.(\ref{Fig:Size1}) we sketch how a small instanton in bulk grow
%to a large Skyrmion on the boundary by holography. This point will
%be discussed further below.
%$Z_c$ is the baryon size on the boundary.
%It is of order $(N_c\lambda)^0$ and large in units of $M_{KK}$. This point
%is important. While the instanton size in bulk is of the
%order of the string length and thus small as $1/\sqrt{\lambda}$
%in units of $M_{KK}$, its image on the boundary is a large Skyrmion.
%
%In~Fig.(\ref{Fig:Size1}) we skech how a small instanton in bulk grow
%to a large Skyrmion on the boundary by holography. This point will
%be discussed further below.
%
%\begin{figure}[]
 % \begin{center}
 %  \includegraphics[width=8cm]{Size1.eps}
 % \caption{Instanton size vs Z}
 % \label{Fig:Size1}
 % \end{center}
%\end{figure}
%
%This observation means that the holographic $1/\sqrt{\lambda}$
%expansion in bulk is matched by a gradient  expansion in
%the boundary. \footnote{ We note that $\mu=d\epsilon/dn_B$
%diverges as $n_B\rightarrow 0$ at order $\lambda^0$ owing to the
%strong attraction in holographic matter.}

%\footnote{The use of a singular instanton through the substitution
%\begin{eqnarray}
%&& \wt{\Pi}(Z) \equiv \Pi(\infty)\left[ \tanh(Z/\wt{\calr})
%\left(1+\half\mathrm{sech}^2(Z/\wt{\calr})\right) -
%\mathrm{sgn}(Z/\wt{\calr})\right] \ ,
%\end{eqnarray} results in a finite $b' \sim 0.08\ \mathrm{MeV}^{-2}$
%with an $n_B^{5/3}$ behavior.
%However, this configuration upsets the divergenceless character of
%the topological current.}.

The various density contributions in (\ref{eos}) can be understood
from the zero density and finite instanton calculation discussed
by Sakai and Sugimoyo to order $N_c\lambda^0$. For that, we
recall that the energy balance for a holographic instanton with
flat $\mathbb{R}^3$ directions reads schematically
as~\cite{Sakai3}

\be
N_c\left( {\bf A}\, \lambda\,\rho^2+ {\bf B}\, \frac 1{\lambda\,\rho^2}\right)
\label{budget}  \ee
leading to an instanton  size in bulk of
order $\rho\sim ({\bf B}/{\bf A})^{1/4}/\sqrt{\lambda}$. The
Coulomb repulsion ${\bf B}$ is $10^4$ times the gravitational
attraction ${\bf A}$ resulting into a size that is of order
$\rho\sim 10/\sqrt{\lambda}$. This parametrically huge repulsion
results in a stiffer equation of state in holographic QCD.
% as we show below.

The linear term in $n_B$ in (\ref{eos}) is just the topological
winding of the $U(N_f)$ flavored instanton in D8 on $S^4$ due to the self duality of the instanton configuration.
It is leading and of order $N_c\lambda$. Geometry is unaffected by
matter. A point-like instanton in bulk corresponds to a very large
Skyrmion on the boundary. The term of order $n_B^{1/3}$ is of
order $N_c\lambda^0$. It corresponds to the {\it attraction} due
to gravity in bulk at finite size. Indeed, the energy of this
term is of order $\lambda\rho^2=\wt{\calr}^2$, as in
(\ref{budget}) favoring smaller and smaller instanton. The energy
per volume for this term is of order $1/\wt{\calr}$. Since in
matter the cell size is of the order of the interparticle distance
$1/n_B^{1/3}$, the $n_B^{1/3}$ follows. The term of order $n_B^2$
is also of order $N_c\lambda^0$. It stems from the Coulomb
repulsion in bulk which is of order
$1/{\lambda\rho^2}=1/\wt{\calr}^2$ since the instanton is static
in 4-space (space-plus-conformal). This contribution is {\it
repulsive} and favors larger size instanton. The corresponding
energy per cell is of order $(Z_c/\wt{\calr})(1/\wt{\calr}^5)$,
since the warping in the conformal direction is subleading in
$1/\lambda$. The $n_B^2$ contribution follows.

For a Skyrmion with a size $Z_c\ll \wt{\calr}$, (\ref{eos})
describes the low density regime. In this regime the use of the
$S^3\times \mathbb{R}$ instanton is likely to give higher energy
than a localized but flat instanton at the pole of $S^3$ say.
Dilute holographic matter is made out of flat
$\mathbb{R}^3$ instantons with (\ref{eos}) providing an upper
bound on the energy per unit volume. This phase breaks
spontaneously chiral symmetry. In the point particle limit,
the equation of state at low density was discussed in~\cite{KSZ2}
\begin{eqnarray}
\e_{\mathrm{p}} \sim N_c \frac{27 \pi^4}{4 \Mkk^2} n_B^2
\label{LOW}
\end{eqnarray}
for low densities after re-scaling $\sqrt{\l}n_B/\l^{3/2}\ra n_B$~\cite{KSZ2}.
The point-like and flat space instanton contribution (\ref{LOW}) at low
density is lower in energy than (\ref{eos}) and therefore favored. This will
be made more explicit below. The $n_B^{1/3}$ is absent in the point like
limit (finite size effect).

As the density is increased (or equivalently
 as $\wt{\calr}$ approaches down to  $Z_c$),  there is a change in the
equation of state (\ref{eos}). For $Z_c=\wt{\calr}$,
\begin{eqnarray}
\varepsilon = M_0 \left[n_B + \frac{a}{\l}n_B^{1/3} + \frac{b'}{\l}
n_B^{5/3} \right] \ ,
\label{Deos}
\end{eqnarray}
with $b$ changing to $b'$
\begin{eqnarray}
b'\equiv \frac{3^6(2\pi^2)^{5/3}}{2^3 \Mkk^2} \,\,.
\end{eqnarray}
The softening of the equation of state at higher density from
$n_B^2$ to $n_B^{5/3}$ follows from a transition from a dilute
gas/liquid  phase to a dense solid/crystal phase. This transition effectively
restores chiral symmetry as we will show later. An estimate of the chiral transition
density follows by comparing the $n_B^2$ term from (\ref{LOW})
to the leading $n_B^{5/3}$ in (\ref{Deos})
\begin{eqnarray}
\e_{\mathrm{s}} \sim N_c \frac{27  \pi^{7/3}}{2^{4/3} \Mkk} n_B^{5/3}
\end{eqnarray}
 By setting $\e_p = \e_s$, the critical transition density follows
\begin{eqnarray}
n_B^c = \frac{4 \Mkk^3}{\pi^5} \,\,.
\end{eqnarray}

\section{Numbers}

To give some estimates of the numbers emerging from the current
discussion, we first recall that in holographic QCD the mass of
one baryon at next to leading order is not unique. We refer to
~\cite{Sakai3} for a more thorough discussion. In particular,
the baryon mass to order $N_c\l^0$ is
\begin{eqnarray}
M_B = M_0\left(1+ \frac{c}{\l}\right) \ ,
\label{next}
\end{eqnarray}
where $c = 27\pi\sqrt{{2}/{15}}$. Thus the interaction energy
per unit volume for the dilute case is
\begin{eqnarray}
E_{\mathrm{int}}^{\mathrm{Dilute}} \equiv \varepsilon - n_B M_B =
\frac{M_0}{\l}\left(a n_B^{1/3} -c n_B  + b\,Z_c\,n_B^{2}\right)
\label{eos1} \ ,
\end{eqnarray}
while for the denser case it is ,
\begin{eqnarray}
E_{\mathrm{int}}^{\mathrm{Dense}} \equiv \varepsilon - n_B M_B =
\frac{M_0}{\l}\left(a n_B^{1/3} -c n_B  + b'\,n_B^{5/3}\right)
\label{Deos1} \ .
\end{eqnarray}
For numerical estimates, we use $\Mkk=500$ MeV and $M_0=940$ MeV
for $N_c=3$~\cite{Sakai3}. Our parameters are
\begin{eqnarray}
\l \sim 53.2 \ , \quad a \sim 0.095\, \mathrm{fm}^{-2}  \ ,
\quad b \sim  2172 \, \mathrm{fm}^3 \ ,  \quad b' \sim  2039 \,
\mathrm{fm}^2 \ ,  \quad c \sim 31 \ . \nn
\end{eqnarray}
The interaction energies are then

\begin{eqnarray}
&& S^3\,\,: E_{\mathrm{int}}^{\mathrm{Dilute}}(\mathrm{GeV}\,
\mathrm{fm}^{-3}) = 0.00168 n_B^{1/3} -0.548
n_B +192 n_B^{2} \ , \label{eos2} \\
&& S^3\,\,: E_{\mathrm{int}}^{\mathrm{Dense}}(\mathrm{GeV}\,
\mathrm{fm}^{-3}) = 0.00168 n_B^{1/3} -0.548 n_B +
36.0\,n_B^{5/3} \ , \\
&& R^3\,\,: E_{\mathrm{int}}^{\mathrm{Dilute,P}}(\mathrm{GeV}\,
\mathrm{fm}^{-3}) = \qquad\qquad -0.548
n_B +60.3\,n_B^{2} \
\end{eqnarray}
where we used $Z_c = 5$.
Notice that due to the smallness of the coefficients of the first two terms, the last term is dominant
even in the relatively small baryon density if it is not much smaller than  $10^{-2}$fm$^{-3}$.

\begin{figure}[]
  \begin{center}
   \includegraphics[width=8cm]{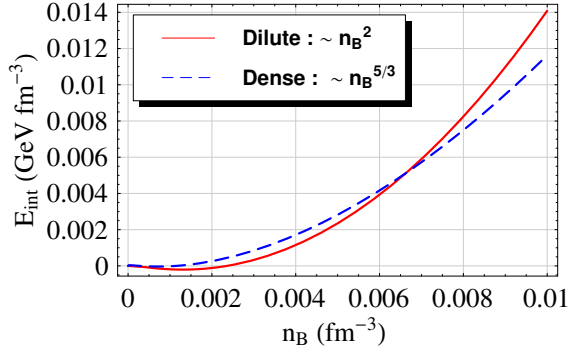}
  \caption{The energy per unit volume: $Z_c = 5$ (red) and $Z_c = \wt{\calr}$ (blue). See text.}
  \label{Fig:Phase}
  \end{center}
\end{figure}

\begin{figure}[]
  \begin{center}
   \includegraphics[width=8cm]{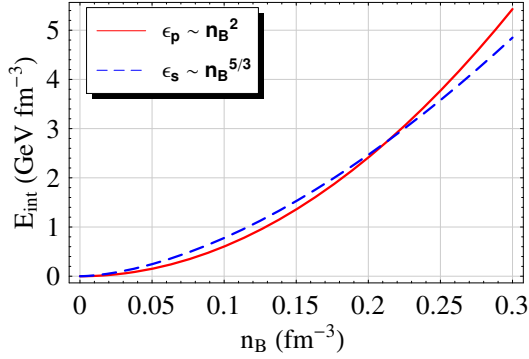}
  \caption{The energy per unit volume:  $Z_c =\infty$ (red) and $Z_c = \wt{\calr}$ (blue). See text.}
  \label{Fig:Phase1}
  \end{center}
\end{figure}

The results on $S^3$:
$E_{\mathrm{int}}^{\mathrm{Dilute}} $ and $
E_{\mathrm{int}}^{\mathrm{Dense}}$ are compared
in~Fig.(\ref{Fig:Phase}) for $Z_c=5$ ( $Z_c=5/\Mkk=1\,{\rm fm}$
with restored dimensions). The results on $S^3$ and $R^3$ are
compared in Fig.~(\ref{Fig:Phase1}). In $R^3$ the instantons are
point-like   or $Z_c\sim \infty$ \cite{KSZ2}. The crossing from $R^3$
to $S^3$ occurs at relatively small densities $n_B^c\approx
1.26\,n_0$ with $n_0=0.17\,{\rm fm}^{-1/3}$ the nuclear matter
density.

\section{Equation of State in General}

The approximation of $T^3$ by $S^3$ suggested at the beginning of
the paper was justified in way in the dilute limit or for small
densities. Phenomenologically, we have found that the chiral
phase transition from $R^4$ to $S^3\times R$ occurs at few times
nuclear matter density in holographic QCD, which is reasonable.
The small size instantons dominate dense matter. This means that
higher order corrections to both the DBI action and the starting
D4 metric are important. While we do not know how to assess them,
we now suggest that they may conspire to be small. Indeed, if we
were not to expand the DBI plus CS actions, that is if we were to
include only these class of higher order corrections our
numerical results change only mildly.

Consider, the total (DBI + CS) action is written as
\begin{eqnarray}
S = -dN_c\l \int d^4 x dZ   \sqrt{A - \inv{\l} B(\Az')^2 } +
\wt{c}N_c \int d^4 x dZ  J \Az \ , \label{Summary.action}
\end{eqnarray}
where
\begin{eqnarray}
&& A \equiv K_\l^{4/3} + \frac{3 b}{\Mkk^2 \Ukk^4} K_\l^{1/3}J +
\frac{3b}{\Ukk^2} K_\l^{5/3}J + \frac{9
b^2}{\Mkk^2\Ukk^6}K_\l^{2/3}J^2
\ , \\
&& B \equiv b K_\l^{5/3} + \frac{3 b^2}{\Mkk^2 \Ukk^4} K_\l^{2/3}J
\ , \qquad  J = \frac{\mathrm{sech}^4(Z/\wt{\calr})} {4
\wt{\calr}^4 } \ ,
\nn \\
&& b\equiv \frac{3^6 \pi^2}{4
 \Mkk^2} \ , \qquad  \wt{c} \equiv \frac{3 }{2\pi^2 \Ukk^3} \ , \qquad d
= \frac{2 \Mkk^4}{3^9 \pi^5}\ .
\end{eqnarray}
The equation of motion is
\begin{eqnarray}
\wt{\Pi}' =  \wt{c}N_c J \ , \label{Gauss.law}
\end{eqnarray}
with
\begin{eqnarray}
\wt{\Pi} \equiv \frac{\dell \call}{\dell\Az'} =  \frac{d N_c B
\Az'}{\sqrt{A-\inv{\l}B(\Az')^2}} \ .
\end{eqnarray}
The integral of of motion is
\begin{eqnarray}
&&\wt{\Pi}(Z) = \wt{\Pi}(\infty)\left[ \tanh(Z/\wt{\calr})
\left(1+\half\mathrm{sech}^2(Z/\wt{\calr})\right)\right]
 \ , \nn \\
&& \wt{\Pi}(\infty) \equiv \frac{\wt{c}N_c}{6\wt{\calr}^3} =
\frac{N_c}{4\pi^2 (\Ukk {\wt\calr})^3} = \sqrt{b}\Pi(\infty) \ .
\end{eqnarray}

\begin{figure}[]
  \begin{center}
   \includegraphics[width=8cm]{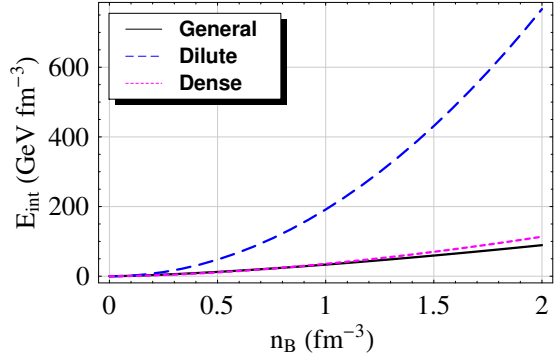}
  \caption{The energy per unit volume.}
  \label{Fig:Eint1}
  \end{center}
\end{figure}

The energy per cell is
\begin{eqnarray}
\cale_{\mathrm{cell}} &=& - \int \e_3dZ ( \call_{\mathrm{DBI}}+
\call_{\mathrm{CS}})  \nn
\\
&=& d N_c \l \int \e_3\  d Z \ \sqrt{\frac{A}{B+\frac{\wt{\Pi}^2}{\l d^2 N_c^2}}} - \int \e_3\ dZ {\cala}_0 \wt{\Pi}'  \ \nn \\
&=& d N_c \l \int \e_3\  d Z \ \sqrt{A + \frac{A\wt{\Pi}^2}{\l
N_c^2 d^2   B}} - \int \e_3\ {\cala}_0 (Z)
\wt{\Pi}(Z)\Big|_{-\infty}^{\infty} \ .
\end{eqnarray}
The energy density ($\varepsilon$) of the crystalline
structure is then
\begin{eqnarray}
\varepsilon &\equiv& \frac{N \cale_{\mathrm{cell}} }{V} \approx
\frac{\cale_{\mathrm{cell}}}{ \int \e_3 } \nn \\
&=& d N_c \l \int   d Z \ \sqrt{A + \frac{A\wt{\Pi}^2}{\l N_c^2
d^2 B}} -  {\cala}_0 (Z) \wt{\Pi}(Z)\Big|_{-\infty}^{\infty} \ ,
\end{eqnarray}
where $N$ is the total number of baryons (cells) and $V$ is the
total volume which is approximated by $N \int \e_3$. The second
term in (\ref{omega}) is
\begin{eqnarray}
 {\cala}_0 (Z) \wt{\Pi}(Z)\Big|_{-\infty}^{\infty}  = 2 {\cala}_0 (\infty) \wt{\Pi}(\infty)
 =N_c \Az(\infty) \frac{1}{2 \pi^2 (\Ukk\wt{\calr})^3} = \m_B n_B \ ,
\end{eqnarray}
where
\begin{eqnarray}
n_B = \frac{1}{\int \e_3} = \frac{1}{2\pi^2 (\Ukk\wt{\calr})^3} =
\frac{1}{2\pi^2 (\sqrt{\l}\calr)^3} \ , \qquad \m_B \equiv N_c
\Az(\infty) \ .
\end{eqnarray}
Since $n_B$ is fixed we may set $\mu_B = 0$. Then the energy
density is
\begin{eqnarray}
\varepsilon &=& d N_c \l \int   d Z \ \left(\sqrt{A +
\frac{A\wt{\Pi}^2}{\l N_c^2 d^2 B}} - K_\l^{2/3}\right)\
,\label{General.energy}
\end{eqnarray}
where we subtracted the vacuum value. Thus the interaction energy
per unit volume is
\begin{eqnarray}
E_{\mathrm{int}} \equiv \varepsilon - n_B M_B \ .
\end{eqnarray}
In Fig.(\ref{Fig:Eint1}) we show the equation of state for the
expanded and unexpanded actions. As expected the corrections are
of order $1/\lambda$ for a finite and small size instanton. The
unexpanded energy is finite for any size of the instanton due to the gravitational warping
factors which are subleading in $1/\lambda$ after rescaling. The
unexpanded results are similar to the expanded ones in the range
of densities explored as it should.  For extremely small $n_B$,
$E_\mathrm{int} \sim 0.00251 n_B^{1/3}$ , however for reasonably large density $n_B \sim 1
\mathrm{fm}^{-3}$, $E_\mathrm{int} \sim 33.9\, n_B^{5/3}$. This
power is consistent and expected from the expansion in eq.(\ref{Deos}).

All general expressions in this section are consistent with the results
quoted above to order $\calo(\l^0)$. Indeed, if for simplicity we set
$\Ukk = \Mkk^{-1}$ with $\Mkk=1$, then $A$ and $B$ reduce to
\begin{eqnarray}
&& A = (1+3bJ)^2 + \inv{\l}\frac{2Z^2}{3}(1+3bJ)(2+3bJ) + \calo(\l^{-2}) \ , \nn \\
&& B = b(1+3bJ) + \calo(\l^{-1}) \ ,
\end{eqnarray}
For example, by considering $3bJ = \wt{F}^2$ and $\wt{\Pi} =
\sqrt{b}\Pi $, we can readily show that (\ref{General.energy}) reduces
to (\ref{Energy})
\begin{eqnarray}
\varepsilon &=&  d N_c \int d Z \ \left[\l {\wt{F}}^2 +
\frac{Z^2}{3}{\wt{F}}^2 + \frac{1}{2}\frac{\Pi^2}{(dN_c)^2}
\right] \ . \nn
\end{eqnarray}

\section{Holographic Skyrmions from Instantons}

\begin{figure}[]
  \begin{center}
   \includegraphics[width=8cm]{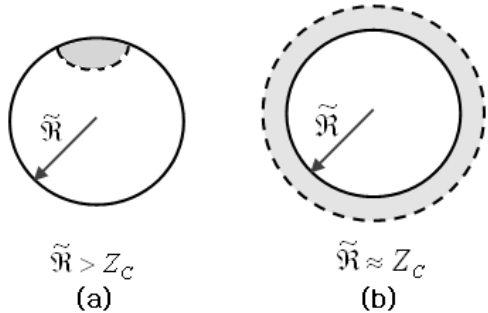}
  \caption{Holographic Skyrmion on $S^3$ on the boundary }
  \label{Fig:Size2}
  \end{center}
\end{figure}

The $S^3\times R$ instanton used in bulk has a very simple Skyrmion
picture on the boundary. From (\ref{INS3}) it follows that the gauge field
at the boundary is $A(\infty,\vec{x}) = U^{-1}dU$. Following~\cite{Sakai3}
we note that $U(\vec{x})$ is just the pion field at the boundary.  When we have a cut-off in $Z$, we replace $A(\infty,\vec{x})$ by $A(Z_c,\vec{x}) $.
$U$ is the boundary Skyrmion field originating from the bulk instanton. Thus $U$
is just the {\it holonomy of the bulk instanton along the conformal direction}:
\be
U(x;Z_c)=P\exp [i \int_0^{Z_c}dZ A^{(instanton)}_Z(Z)]
\ee
When  the density is large  and $Z_c\sim \cal R$,  the instanton has a support covering the whole three sphere, therefore the resulting Skyrmion should be
\begin{eqnarray}
U (\vec{x})\simeq  \sigma(\vec{x}) + i \t_a
\Pi^a(\vec{x}) =e^{i\t_a \hat{r}^a(\theta, \phi)\,\psi}\ ,
\label{SK3}
\end{eqnarray}
which is the identity map as $(\psi ,\theta, \phi)$ are the canonical angles for
the unit $S^3$. The local Jacobian matrix for this map from $S^3$ to $S^3$
is $J^{ai}=\partial\Pi^a/\partial x_i={\bf 1}^{ai}/R$, proportional to the identity.
The baryon density for this map is ${\rm det J}/{\rm vol S^3}=1/(2\pi^2 R^3)$ in
agreement with bulk holography. The scalar field $\sigma(\vec{x})=\cos \psi$ measures
the {\it chiral condensate} and averages to zero on $S^3$
\begin{eqnarray}
\frac{<\overline{q}q>_{S^3}}{<\overline{q}q>_{R^3}}
=<\sigma (x)>_{S^3}=\frac 2{\pi}\int_0^\pi\,d\psi
{\rm sin}^2\psi \,{\rm cos}\psi =0\,\,.\label{chirestore}
\end{eqnarray}
The $S^3\times R$ instanton in (\ref{INS3}) corresponds to a boundary Skyrmion
on $S^3$ with restored chiral symmetry on the average.
We should notice that
the chiral condensation is p-wave over a cell while the density in this case  is approximately   constant over a cell.  But it is certainly not a constant.
In fact this is a result  consistent with
 ref. \cite{Rozali} where it was argued that there can not be an uniform distribution.
In Fig.~\ref{Fig:Size2},
we show schematically how a Skyrmion of size $Z_c$ looks on $S^3$ as a function
of $\calr$. (a) corresponds to the dilute phase with broken chiral symmetry, while
(b) describes the dense phase with restored chiral symmetry.

In previous section, $Z_c$ was introduced as a cut-off of the action  bigger than the
instanton size.  Here we give interpretation of $Z_c$ as the size of the Skyrmion on the boundary. Note that the $R^4$ BPS instanton used
in bulk in~\cite{Sakai3} for the description of a single baryon, yields
a boundary Skyrmion as
\begin{eqnarray}
U(\vec{x})={Z_c}/{\xi_c}-i\vec{\tau}\cdot\vec{x}/\xi_c
\label{SK}
\end{eqnarray}
with $\xi_c^2=Z_c^2+\vec{x}^2+\rho^2$ and
this is the analogue of the unit map (\ref{SK3}) with ${\rm tan}\psi=x/\xi_c$.
Notice that {\it  while the size of the instanton is $\rho$,
the size of the Skyrmion is $\sqrt{Z_c^2+\rho^2}$.}
If $\rho \ll Z_c$, $Z_c$ itself is the size of the Skyrmion,  hence our interpretation above comes. Holography transmutes a small size instanton $\rho$ in bulk to a large size Skyrmion on the boundary.

At small densities with $\wt\calr \gg Z_c$, one can replace the spherical cell
 by a flat space and  the map (\ref{SK})
is relevant, while at high density  $\wt\calr \leq Z_c$ the map
(\ref{SK3}) is relevant. On $S^3$ this is pictorially depicted in
Fig.(\ref{Fig:Size2}).
Notice also that (a) has broken chiral symmetry while (b)
has restored chiral symmetry effectively( See eq. (\ref{chirestore})) .
Again, in this case,  our $S^3\times R$ instanton
in bulk describes the high density phase in holographic QCD with
restored chiral symmetry. At low densities the energy density is
about $n_B^2$ as discussed by many~\cite{Finitemu} in qualitative
agreement with our figure here. The $n_B^2$ term is sourced by
Coulomb's repulsion in both cases. The description on $S^3$
carries larger energy density than on $R^3$ and is therefore
unfavorable energitically. It is favorable at higher densities.
The transition occurs at about $\calr=Z_c$, or $n_B^c=1/(2\pi^2\,
Z_c)$, resulting into an energy density of $n_B^{5/3}$. The value
of $n_B^c$ was estimated above.

The determination of $Z_c$ or equivalently the critical size of $\wt\calr$ depends
on the energetics of the SS model. It is worth pointing that the single baryon mass
analysis on $S^3$ as discussed in~\cite{DENSES3} allows a considerable simplification
of this issue when the Skyrme model is used. We now note that this is justified in
holographic QCD as small size instantons in bulk with $\rho=\wt\calr/\sqrt{\lambda}$
map onto a large size Skyrmion on the boundary with $Z_c\gg \rho$. So the small size
instanton expansion in bulk maps onto the gradient expansion in $1/Z_c$ on the boundary.
Limiting the SS model on the boundary to the Skyrme model with $f_\pi$ and $e_S$ fixed by
holography yields the specifics of the Skyrmion on the boundary to order $\lambda$.

\begin{figure}[]
  \begin{center}
   \includegraphics[width=10cm]{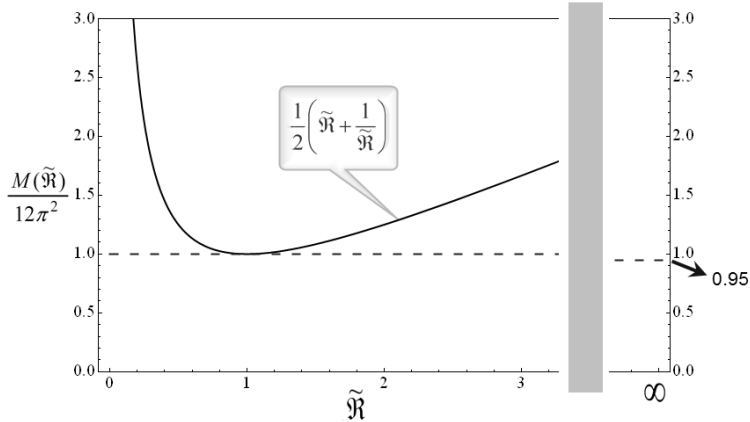}
  \caption{Holographic Skyrmion mass on $S^3$: order $\lambda$}
  \label{MASS}
  \end{center}
\end{figure}

\begin{figure}[]
  \begin{center}
   \includegraphics[width=10cm]{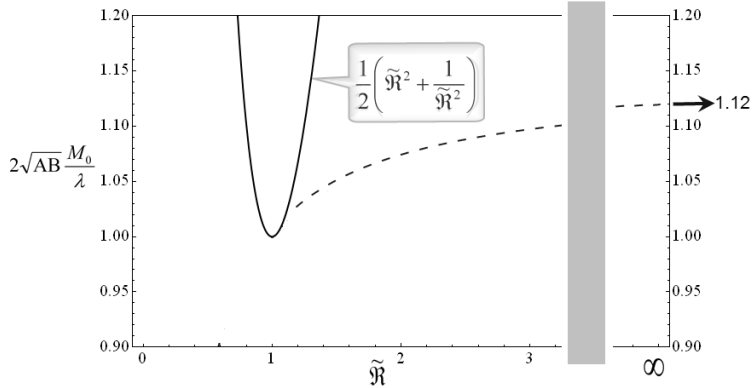}
  \caption{Holographic Skyrmion mass on $S^3$: order $\lambda^0$}
  \label{MASS2}
  \end{center}
\end{figure}

In Fig.~(\ref{MASS}) we show how the holgraphic Skyrmion mass on
$S^3$ to order $\lambda$ changes with $\wt\calr$ the radius of
$S^3$ following~\cite{DENSES3}. The units of mass and length are
respectively~\cite{Sakai3}
\begin{eqnarray}
\frac{f_\pi}{2\sqrt{2}e_S}&=&(\lambda N_c)\,\Mkk \,\frac{\sqrt{{\bf b}/2\pi}}{54\pi^5}\\
\frac{\sqrt{2}}{e_S f_\pi}&=& (1/\Mkk)\,{\sqrt{8\bf b}}/{\pi^3}\,\,.\nonumber
\end{eqnarray}
with ${\bf b}=15.25$ and $L={\calr}$. We note that the mass
$M_0=8\pi^2\kappa \Mkk$ corresponds to the point $0.95$ at
$\wt\calr=\infty$ which matches the unit map result as expected.
In Fig.~(\ref{MASS2}) we show the same curve to order $1/\lambda$.
Here the energetics is determined in bulk as the chiral
Lagrangian in the SS model is not known beyond the order
$\lambda$. Specifically,
\begin{eqnarray}
\frac{M_0}{\l}\left({\bf A}\wt{\calr}^2+\frac{\bf
B}{\wt{\calr}^2}\right)
\end{eqnarray}
with ${\bf A}= (\pi^2 - 6)/36 \sim 0.11$ and ${\bf B}= (3^6
\pi^2)/4 \sim 1799 $. The units of mass and lengths are
\begin{eqnarray}
\left({\bf B}/{\bf A}\right)^{1/4} &\sim& 11.4\nonumber\\
2\sqrt{{\bf AB}} &\sim& 27.8
\end{eqnarray}
The point $1.12$ is the $1/\lambda$ corrected mass (\ref{next}) in these units.
Finally, it is interesting to note that the holographic Skyrme model on $S^3$ yields
{\it naively} the following equation of state
\begin{eqnarray}
\varepsilon =  M_0 (n_B + a_S\, n_B^{2/3} + b_S\,
n_B^{4/3} ) \ ,
\label{ES.density}
\end{eqnarray}
as first noted in the context of the canonical Skyrme model~\cite{KPR}. The $n_B^{2/3}$
for the Skyrmion stems from the universal current algebra $(\nabla\Pi)^2$ term which
is attractive and scales as $1/\rho^2$ as opposed to $1/\rho$ from the finite size
instanton in bulk. The $n_B^{4/3}$ for the Skyrmion stems from the repulsive
Coulomb contribution per unit 3-volume $(1/\rho)/\rho^3$ from the Skyrme term as opposed to
the repulsive Coulomb contribution  per unit 3-volume $(1/\rho^2)/\rho^3$ in the instanton in bulk.
We recall that Coulomb's law in 1+D dimensions is $1/\rho^{D-2}$.

At high density the {\it naive} scalings in (\ref{ES.density}) obtained at the boundary differs
from (\ref{Deos}) obtained in bulk in two essential ways: i)
$a_S$ and $b_S$ are of order $N_c^0\lambda^0$ on the boundary while
their bulk contributions are of order $N_c^0/\lambda$; ii) the scaling
with $n_B$ appears to differ by an extra (spatial) dimension, $D=3$ on the
boundary and $D=4$ in bulk. These differences can be understood
by noting that the size of the holographic Skyrmion is $Z_c$.
This means that the chiral gradients $L_i=U^{-1}\partial_iU$ are nearly zero on the boundary
with $U\sim {\bf 1}$,
except on an the shell $|\vec{x}|\approx Z_c$ of thickness $1/\sqrt{\lambda}$
to ensure that the topological baryon charge is finite
~\footnote{We note that for $U\sim {\bf 1}$ the Skyrmion obeys the Faddeev-Bogomolnyi bound
since the classical equations of motion are fulfilled.}. This renders $a_S$ and $b_S$
in (\ref{ES.density}) effectively of order $1/\lambda$ as noted in bulk.
%Also, since
%$Z_c$ is large but {\it not fixed} quantum mechanically, it means that we need to average over
%the large sizes $Z_c$ at the boundary. Litterally, this means that the holographic Skyrmion through its
%large size leaks in the conformal direction making its space effectively $D=4$ dimensional.
%Holographic QCD provides a specific measure for the size along the conformal direction in bulk.

\section{Comparison with Nuclear Models}

\begin{figure}[]
  \begin{center}
   \includegraphics[width=10cm]{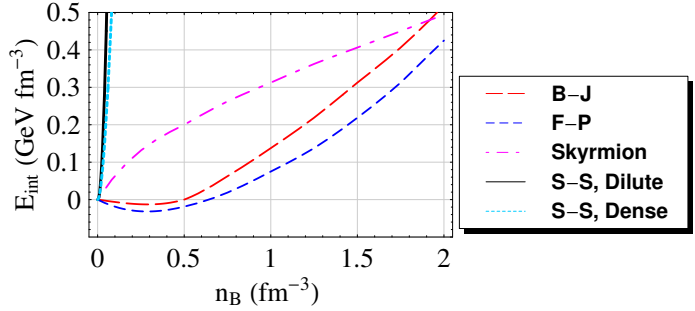}
  \caption{The energy per unit volume as a function of baryon density,
   for pure skyrmions, for the calculations of Bethe and Johnson
   \cite{KPR,Bethe}
   and of Friedman and Pandharipande \cite{KPR,Friedman}, and for our
   instanton model based on Sakai-Sugimoto model for dilute and dense case.}
  \label{Fig:Eint}
  \end{center}
\end{figure}

In Fig.(\ref{Fig:Eint}) we compare the interaction energy
(\ref{eos1}) and (\ref{Deos1}) with other hadronic models
including Skyrme's chiral model. Holographic matter is
subtantially stiffer as explained through the energy budget in
(\ref{budget}). The reason can be traced back to the fact that
for a {\it single} baryon the repulsion already dwarfs the
attraction in holographic QCD

At high densities $\varepsilon$ in (\ref{Deos1}) is approximated as
\begin{eqnarray}
\varepsilon \ \sim\ \frac{N_c 3^3 (2\pi^2)^{5/3}  }{2^3 \pi \Mkk
}\, n_B^{5/3} \quad \sim \quad  36  n_B^{5/3}\
(\mathrm{GeV}\,\mathrm{fm}^{-3}) \quad \mathrm{for}\, N_c = 3 \ ,
\end{eqnarray}
and whatever $N_f$ since the flavoured instanton in bulk is
always $2\times 2$.
This behaviour is different from that of free massless quarks in
$D=3$ ($\varepsilon_3$) but similar to {\it massive} quarks in
$D=3$ ($\varepsilon_3'$). Specifically,
\begin{eqnarray}
&&\varepsilon_3 = \frac{N_c}{N_f^{1/3}} \frac{3^{4/3}
\pi^{2/3}}{4} n_B^{4/3}  \quad \sim \quad  5.52
 \, n_B^{4/3}(\mathrm{GeV}\,\mathrm{fm}^{-3}) \quad
\mathrm{for}\, N_c = 3, N_f =2 \ , \\
&&\varepsilon_3' = \frac{N_c}{N_f^{2/3}}\, \frac{3^{3/5}\pi^{4/3}}{10} \frac{1}{m}n_B^{5/3}
 \quad \sim \quad 1.68\inv{m}\, n_B^{5/3}(\mathrm{GeV}\,\mathrm{fm}^{-3}) \quad \mathrm{for}\, N_c = 3, N_f
 =2\ .
\end{eqnarray}
So at strong coupling
\begin{eqnarray}
&&\frac{\varepsilon}{\varepsilon_3'} = N_f^{2/3}\left(\frac{9^{6/5}5}{2^{1/3}}\right)
 \frac{m}{\Mkk}\quad \sim\quad  \frac{88m}{\Mkk}  \quad \mathrm{for}\, N_f
 =2\ ,
\end{eqnarray}
independently of $\l$ and $N_c$.
As chiral symmetry is restored in the high density phase, the
comparison to the the free massive quark phase in $D=3$ suggests
that the mass $m\sim \Mkk/88$ is a chirally symmetric screening mass.
While the chiral transition restores chiral symmetry it still confines
baryons.

\section{Conclusions}

We have provided a holographic description of dense and cold
hadronic matter using the brane model put forward by Sakai
and Sugimoto~\cite{Sakai3}. At large $N_c$ the matter crystallizes
and can be treated in the Wigner-Seitz approximation on $T^3$. For simplicity,
the Wigner-Seitz cell was further approximated by $S^3$ in space leading
to a simple instanton configuration on $S^3\times R$ with $R$ the conformal
space. The resulting equation of state at next to leading order in $\lambda$
shows a free quark behavior at high density, although the overall coefficient
is cutoff sensitive and large resulting into a stiff euation of state.

At high densities the gauge gradients are
of order $\sqrt{\lambda}$ so the DBI action may not be enough to fix the brane
dynamics at order $N_c\lambda^0$~\cite{Sakai3}. Also our simplification of $T^3$
by $S^3$ while justified at low density, involves curvature corrections at high
densities. However, we believe that the essentials of dense matter in holographic
QCD are already exposed on $S^3$ with a small  attraction leading
$n_B^{1/3}$ and a large Coulomb repulsion leading $n_B^{5/3}$, where 5/3 is the power of non-relativistic fermion.
It is interesting to notice that the coulomb interaction in the bulk counts the fermi statistics
in the boundary. The repulsion
is $10^4$ times the attraction resulting into a very stiff equation of state.
Changing $S^3$ to $T^3$ will not affect the outcome quantitativaly we believe.
Indeed, this is the case for dense Skyrmions~\cite{DENSES3}.

The present work expands on the original ideas developed in~\cite{Finitemu}.
Our calculations with finite size instantons are closer to those presented
in  reference  \cite{Rozali} where finite size and homogeneous
instantons were used through a variational estimate in $R^3\times R$.
Their arguments yield $n_B$ instead of the $n_B^{1/3}$ we have reported in the
equation of state at next to leading order with our $S^3\times R$ instanton.

The inhomegeneous $S^3\times R$ description of the crystal
suggests that at high density, chiral symmetry is restored on the
average. Indeed, since the dual of the instanton cell is the
Skyrmion cell with a pion field restricted to $S^3$ in space.
High density matter corresponds to small size $S^3$ where the
pion field becomes just the unit map~\cite{DENSES3}. The
corresponding chiral condensate on $S^3$ is seen to vanish as half
of $S^3$ carries positive chiral condensate, while the other half
carries negative chiral condensate so that on the average the
chiral condensate is zero. This restoration of chiral symmetry is
due to the formation of the crystal in the spatial direction in
holographic QCD even though the $\De$-$\DeB$ configuration is still
attached. In other words, the left and right D8 branes cease to
talk to each other through the spatial directions not the
conformal direction when they {\it crystallize at large $N_c$}.

The present crystal analysis is classical in bulk. A quantum analysis
including vibrational and rotational motion is needed. These corrections
are subleading in $1/N_c$ and should be estimated for a more thorough
phenomenological discussion. Also, the inhomogeneous phase can be probed
approximatly by a dilute gas of instantons on $T^3$ allowing for a lower
energetics than on $S^3$. These issues and others will be discussed elsewhere.

\section{Acknowledgments}
The work of KYK and IZ was supported in part by US-DOE grants
DE-FG02-88ER40388 and DE-FG03-97ER4014. The work of SJS was
supported by KOSEF Grant  R01-2007-000-10214-0  and the SRC Program
of the KOSEF through the CQUEST  grant R11-2005-021.

\end{document}